\documentclass[10pt]{article}
\usepackage[letterpaper,margin=0.91in]{geometry}
\usepackage[T1]{fontenc}
\usepackage[utf8]{inputenc}
\usepackage{lmodern}
\usepackage{amsmath,amssymb,bm,booktabs,graphicx,microtype,siunitx,hyperref,xcolor}
\usepackage[labelfont=bf,font=small]{caption}
\usepackage{placeins}
\hypersetup{colorlinks=true,linkcolor=blue!50!black,citecolor=blue!50!black,urlcolor=blue!50!black}
\title{Reliability of Fisher-information bounds for dielectric metalenses: a site-resolved RCWA audit}
\author{Juan Sumaya-Mart\'inez$^{1,*}$ and Omar Olmos-L\'opez$^{2}$\\[3pt]
\small $^{1}$Faculty of Sciences, Universidad Aut\'onoma del Estado de M\'exico, Toluca 50000, M\'exico\\
\small $^{2}$Tecnol\'ogico de Monterrey, Campus Toluca, Estado de M\'exico, M\'exico\\
\small $^{*}$Corresponding author: jsm@uaemex.mx; ORCID (J.S.-M.): 0000-0002-7032-8824}
\date{}
\begin{document}
\maketitle
\begin{abstract}
We assess the reliability of localization precision predicted for a 1,716-nanopillar dielectric metalens using a site-resolved complex RCWA library and a vectorial Poisson Fisher model. An independent calculation reproduces the archived local axial Cram\'er--Rao bound of 20.7738 nm to within 0.0044 nm, yielding 20.7695 nm under the same ideal detector and 3,000 detected signal photons. High-order periodic-cell calculations reveal a 0.1335-rad phase difference at a selected diameter; different substrate-index assumptions prevent attributing this difference solely to modal truncation. Replacing the recalculated coefficients at 39 actual sites changes the bound to 20.7522 nm. Restricted phase-mask benchmarks and pixel-integration checks further distinguish a stable statistical prediction from electromagnetic convergence. We provide a reproducible, model-conditional benchmark, not a physical-device performance claim.
\end{abstract}
\noindent\textbf{Keywords:} dielectric metalenses; rigorous coupled-wave analysis; vectorial imaging; photon-limited localization; Fisher information; numerical convergence; reproducible simulation.

\section{Introduction}
A narrow image is not automatically a good position measurement. In the photon-limited regime, the precision with which a point source can be localized is controlled by changes in the distribution of expected photon counts as the source moves. This distinction motivates engineered point-spread functions (PSFs), for which the image shape can be deliberately altered to make axial motion observable even when conventional focusing would provide little depth discrimination. Fisher-information-based PSF optimization is established, with both theoretical and experimental milestones in single-emitter imaging \cite{shechtman2014,shechtman2015}. Double-helix PSFs and related approaches demonstrate how axial displacement can be encoded into changes in image structure \cite{pavani2009,grover2012,wang2023,zhang2026}. Recent PSF calibration and black-box-lens modeling studies also show why an apparently informative image can produce biased localization if the assumed forward model is wrong \cite{petrov2017,xiao2025}.

Dielectric meta-optics provides a compact route to tailored wavefronts \cite{yu2014,khor2016}. Yet engineered PSFs have also been produced using dielectric metalenses, so that ability alone is not a new contribution \cite{shanei2017}. What matters for a statistically optimized metalens is whether a useful PSF survives the change from an ideal pupil to a finite set of physical meta-atoms. Diameter-dependent amplitude, phase, polarization response, angular incidence, and coupling between nearby pillars can all modify the predicted localization precision. A comparison based only on the brightest focal spot or on the throughput of an isolated cell leaves the inference problem unresolved.

A previous local-periodic model of a TiO$_2$ metalens predicted an axial Cram\'er--Rao standard-deviation bound of 20.77 nm, compared with 27.49 nm for a particular astigmatic reference at 3000 detected signal photons and 150 background photons. We now \emph{independently reproduce} the original axial bound from the recovered site-by-site diameter map and complex transmission lookup, while preserving its original statistical assumptions. The historical astigmatic reference was optimized under different settings from the matched-family screens presented below and cannot be used as their direct baseline. We also ask a complementary numerical question: \emph{how stable is the route from a finite RCWA transmission library to calculated PSFs and localization Fisher information?}

Our contribution is a three-stage reproducibility and sensitivity audit rather than a new experimentally certified metalens: (i) reconstruct the exact site-resolved local-periodic prediction instead of a schematic phase-only pupil, (ii) perturb the actual nanopillar assignment with independently recalculated periodic-cell coefficients at identified sites, and (iii) demonstrate that stable Fisher outputs, sampled-pupil convergence and complex-electromagnetic convergence are logically distinct. This makes the model's domain of validity quantitatively testable. The result complements optical PSF optimization and does not claim to replace direct finite-device Maxwell calculations.

\begin{table}[t]
\centering\caption{Comparison with representative published approaches. The table states demonstrated scope, not a ranking of experimental performance.}\label{tab:related}
\begin{tabular}{p{.27\linewidth}p{.34\linewidth}p{.27\linewidth}}\toprule
Prior study & Previously demonstrated & What this work assesses \\\midrule
Shechtman et al. \cite{shechtman2014}&Fisher-based PSF optimization and imaging&Numerical transfer to discrete unit-cell response\\
Shanei et al. \cite{shanei2017}&Engineered-PSF dielectric metalenses&Complex cell-phase reliability\\
Xiao et al. \cite{xiao2025}&Calibration and PSF design for imaging systems&Detector-model and Fisher consistency\\
Moreno et al. \cite{moreno2021,delolmo2021}&Quantitative optical prediction with comparison to measurements/models&Explicit measurable numerical error budgets\\\bottomrule
\end{tabular}
\end{table}

This question has both electromagnetic and statistical parts. The unit-cell calculation must stabilize complex transmission, not merely total power. The map from desired pupil phase to a sparse library can introduce discontinuous changes in chosen pillar diameters. Finally, Fisher information depends on positional derivatives of \emph{pixel-integrated} intensities, nuisance parameters, and the detected-photon normalization. We assemble those steps in one reproducible numerical chain and report what it does---and does not---establish. The presentation follows a conventional optical research article: physical model, quantitative validation, results, and limitations; supporting numerical diagnostics are moved to the accompanying supplement.

\section{Physical system and computational scope}
\begin{figure}[t]
\centering
\includegraphics[width=.98\linewidth]{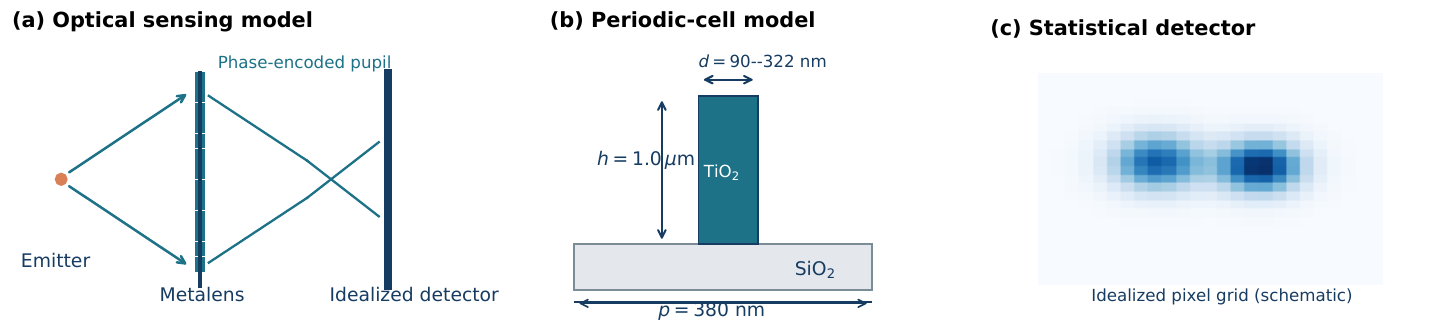}
\caption{\textbf{Conceptual metalens-to-detector model.} (a) A point emitter illuminates an aperture populated by TiO$_2$ pillars on fused silica; (b) the site pitch, pillar height, and available diameter range; (c) a synthetic, schematic pixel grid used only to illustrate the statistical observation model, not a computed or measured PSF. The actual pillar assignment is shown in Fig.~\ref{fig:recovered}.}
\label{fig:system}
\end{figure}

The representative system (Fig.~\ref{fig:system}) uses a vacuum wavelength $\lambda=850$ nm, pitch $p=380$ nm, cylindrical-pillar height $h=1.0~\mu$m, and candidate diameters between 90 and 322 nm. The aperture diameter is 17.86 $\mu$m and the nominal focusing distance is 7.88 $\mu$m; these dimensions correspond to a geometrical numerical aperture near 0.75 in air. TiO$_2$ is modeled with the real refractive index $n=2.4116$ at the chosen wavelength. The recalculated seven-diameter cell model uses a fused-silica substrate index $n_s=1.45$, whereas the recovered historical complex library uses $n_s=1.4585$ in its flux normalization. Both are dispersionless and lossless assumptions, not fitted measurements. Accordingly, archived-to-independent coefficient differences reflect a change in model parameters as well as discretization; only within-solver harmonic sequences are suitable for a clean truncation check.

\begin{table}[t]
\centering\caption{Specified parameters of the reproducible surrogate. The fixed detector here belongs to an independent surrogate model; a separate section reproduces the recovered original design in its archived 48$\times$48 Fourier detector.}\label{tab:params}
\begin{tabular}{ll}\toprule
Vacuum wavelength & 850 nm\\
Nominal NA, aperture diameter & 0.75, 17.86 $\mu$m\\
Nominal focus & 7.88 $\mu$m\\
Square pillar pitch, pillar height & 380 nm, 1.0 $\mu$m\\
Seven \emph{surrogate} pillar diameters & 90--322 nm\\
Detector, pixel pitch & $25\times25$, 0.20 $\mu$m\\
Photon counts & $N_s=3000$, $b=150$\\
Detector quadrature & Gauss--Legendre $3\times3$\\
\bottomrule\end{tabular}
\end{table}

The original design has 113 distinct pillar diameters; the seven-diameter library in Table~\ref{tab:params} refers only to the independent surrogate convergence study. The study separates three distinct forward models. The \emph{periodic-cell Maxwell model} resolves normally incident light scattered by an infinite array of identical cylindrical nanopillars. The \emph{local-periodic pupil model} assigns one of those complex transmission coefficients to each site of a finite aperture, ignoring nonlocal coupling and incidence-angle dependence. The \emph{vectorial detector model} propagates the resulting pupil field, integrates the intensity of each pixel, and computes a Fisher matrix. None of these stages, individually or in combination, constitutes an open-boundary three-dimensional Maxwell solution of the finite metalens illuminated by a vector electric dipole. The distinction is essential to interpreting the results.

\section{Maxwell and statistical forward models}
\subsection{Complex Fourier-modal transmission}
For each periodic unit cell, the permittivity in the transverse plane is expanded over a truncated set of Fourier orders,
\begin{equation}
 \epsilon(x,y)=\sum_{m,n}\epsilon_{mn}\exp\!\left[i\frac{2\pi}{p}(mx+ny)\right].\label{eq:eps}
\end{equation}
Tangential electric and magnetic Fourier amplitudes are propagated through the pillar layer and matched to the substrate and ambient half spaces. The primary outputs are the complex zeroth-order transmission coefficient $t_0(d)$, reflected and transmitted powers $R(d),T(d)$, and polarization-coupling diagnostics. At normal incidence and for the present subwavelength pitch, the only freely propagating far-field diffraction order is the zeroth order. The normalization of $t_0$ is chosen such that the zeroth-order flux is $|\tilde t_0|^2=T$ when there is no other propagating transmitted channel.

Agreement of $R+T$ with unity is a valuable energy-balance check for a lossless unit cell, but it is not an error bound for its transmitted phase. The numerical analysis of discontinuous periodic permittivity is known to require care in stable matrix propagation, Fourier factorization, and eigenproblem construction \cite{moharam1995,lalanne1996,li1996}. We therefore examine separately the power variation, the wrapped phase difference,
\begin{equation}
\delta\phi_d=\arg\!\left[\tilde t_{0,225}(d)\tilde t_{0,169}^{*}(d)\right],\label{eq:phase}
\end{equation}
and the complex discrepancy $|\tilde t_{0,225}-\tilde t_{0,169}|$ as modal truncation is refined. A thin-film Fresnel solution provides a useful independent limiting-case verification, but periodic-cell and full-array convergence remain distinct problems.

\subsection{Pillar assignment and vectorial pupil propagation}
Let $\mathbf{u}=(u,v)$ span the normalized circular pupil and $k=2\pi/\lambda$. For a nominal focal length $f$ and radius $a$, the ideal focusing phase is
\begin{equation}
\phi_f(\mathbf{u})=-k\left(\sqrt{f^2+a^2|\mathbf{u}|^2}-f\right).
\end{equation}
The test families add a residual astigmatic phase $\phi_A=1.5(u^2-v^2)$ or a separate mixed residual phase $\phi_M=\phi_A+0.5(u^3-3uv^2)+0.25uv$. These are \emph{illustrative masks}; the mixed phase is not an optimized Tetrapod or double-helix mask, and neither family is the original Fisher-optimized geometry recovered in Sec.~\ref{sec:recovered}.

For each site $j$ the target phase $\phi_f+\phi_{A,M}$ is mapped to the nearest phase in the seven-point library. The complex coefficient after removing the nominal focusing phase defines the residual pupil response $t_{\mathrm{res},j}$. In an angular-spectrum model, the three detected electric-field components can be written schematically as
\begin{equation}
 E_q(\mathbf{X};\mathbf{r})=\iint_{|\mathbf{u}|\le1} P(\mathbf{u})\,e_q(\mathbf{u})\,t_{\mathrm{res}}(\mathbf{u})
 e^{i[k_x(X-x)+k_y(Y-y)+(k_z-k)z]}\,d^2\mathbf{u},\label{eq:vector}
\end{equation}
where $\mathbf{r}=(x,y,z)$ is a source-displacement parameter in the model, $\mathbf{X}=(X,Y)$ is a detector coordinate, $k_x=k\,\mathrm{NA}\,u$, $k_y=k\,\mathrm{NA}\,v$, and $k_z=(k^2-k_x^2-k_y^2)^{1/2}$. The vector $(e_x,e_y,e_z)$ is a transverse polarization field satisfying $\mathbf{k}\cdot\mathbf{e}=0$ in the homogeneous propagation region. An idealized apodization $P$ is used consistently across the comparisons; the full high-aperture vector-diffraction problem is more involved \cite{richards1959}.

The spatial derivatives are evaluated \emph{before} squaring the field: $\partial_x E_q=-ik_x E_q$ at the integrand level, $\partial_y E_q=-ik_y E_q$, and $\partial_z E_q=i(k_z-k)E_q$. Then
\begin{equation}
 I=\sum_q|E_q|^2,\qquad \partial_a I=2\operatorname{Re}\sum_q E_q^*\partial_a E_q,\label{eq:di}
\end{equation}
for $a=x,y,z$. These identities avoid artificial cubic-interpolation shifts of a sampled PSF. They are identities of the stated model, not a proof that Eq.~\eqref{eq:vector} captures the vector dipole--substrate scattering problem exactly.

\subsection{Pixel integration and nuisance-aware Fisher information}
With the same fixed detector for both masks, intensity is integrated over finite pixels $\Omega_j$,
\begin{equation}
 p_j(\mathbf r)=\frac{\int_{\Omega_j} I(\mathbf X;\mathbf r)\,d^2\mathbf X}{\sum_l\int_{\Omega_l} I(\mathbf X;\mathbf r)\,d^2\mathbf X},\qquad
 \mu_j=N_s p_j+\frac{b}{M}.\label{eq:poisson}
\end{equation}
For independent Poisson photoelectrons and parameter vector $\bm\theta=(x,y,z,N_s,b)$, the Fisher matrix is
\begin{equation}
 F_{ab}=\sum_{j=1}^{M}\frac{1}{\mu_j}\frac{\partial\mu_j}{\partial\theta_a}\frac{\partial\mu_j}{\partial\theta_b}.\label{eq:fisher}
\end{equation}
Partitioning position $\mathbf r$ from the nuisance parameters $\bm\nu=(N_s,b)$ yields the effective positional information
\begin{equation}
 F_{\mathrm{eff}}=F_{rr}-F_{r\nu}F_{\nu\nu}^{-1}F_{\nu r},\qquad
\sigma_z\ge\sqrt{\left[F_{\mathrm{eff}}^{-1}\right]_{zz}}.\label{eq:schur}
\end{equation}
The reported $\sigma_z$ is a local lower bound on an \emph{unbiased} estimator's standard deviation in this forward model, not a directly observed RMSE. Importantly, $N_s$ is fixed at the \emph{detected} photon count: equal incident-photon budgets would require independently validated collection and transmission efficiencies, which are not available for the finite metalens. The normalized-PSF calculation therefore cannot establish an absolute device efficiency advantage. Pixelation and photon statistics are well-known constraints on localization precision \cite{ober2004}.

\section{Reproducing the site-resolved Fisher-optimized metalens and restricted PSF benchmarks}\label{sec:recovered}
\begin{figure}[htbp]
\centering\includegraphics[width=.47\linewidth]{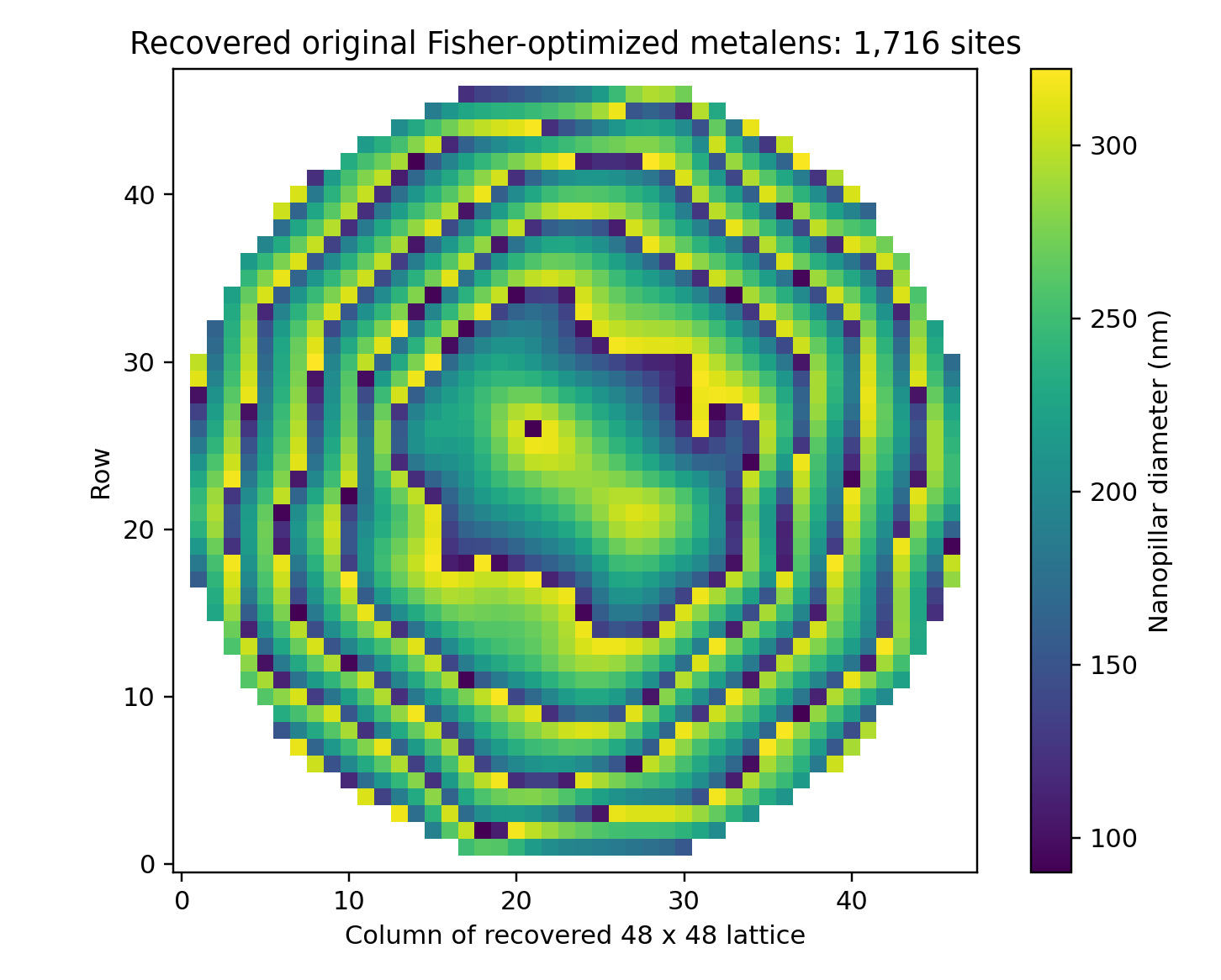}\hfill
\includegraphics[width=.50\linewidth]{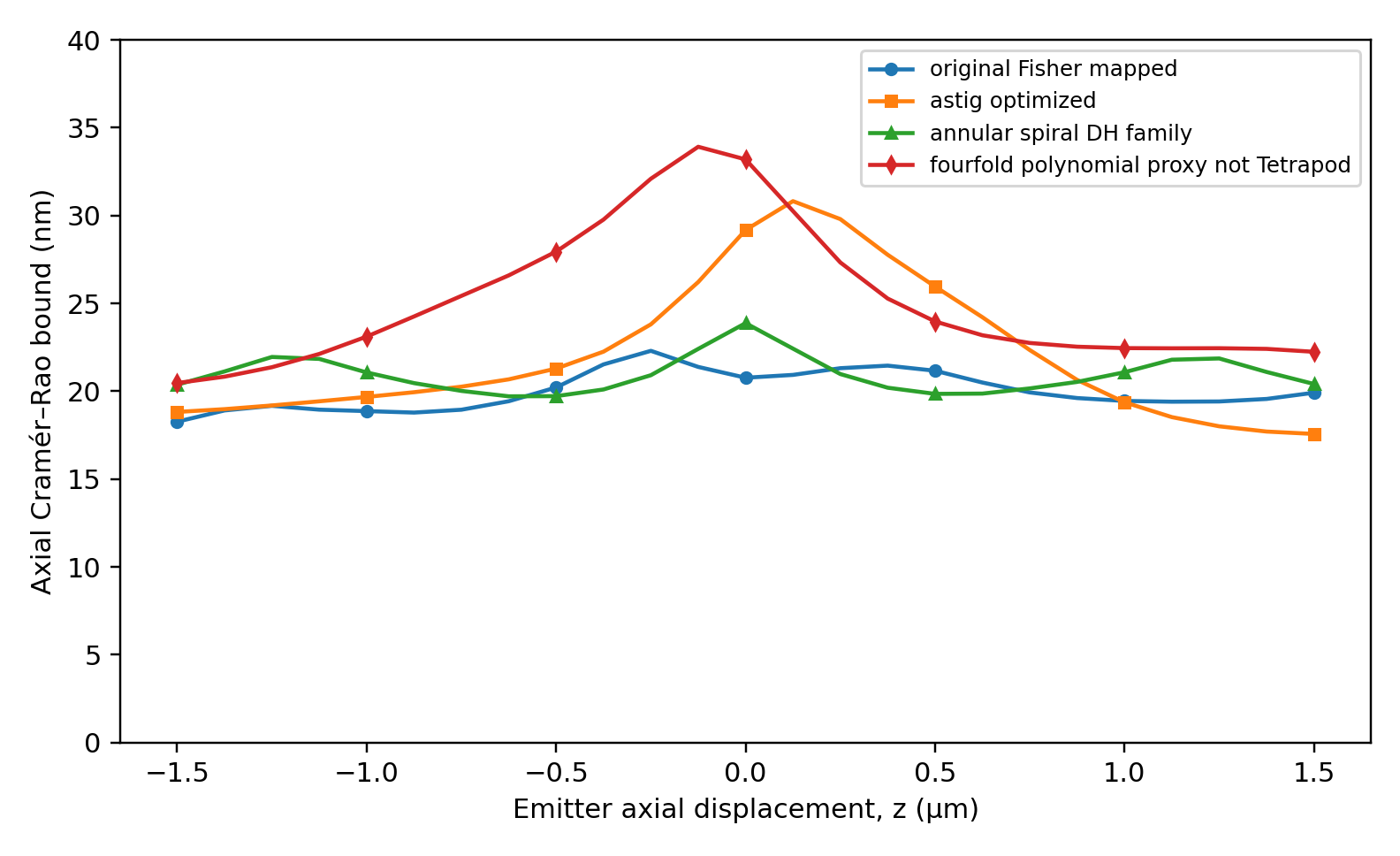}
\caption{\textbf{Recovered metalens and matched local Fisher benchmark.} (a) Actual 48$\times$48 site-resolved nanopillar diameters, with 1,716 active locations and 380-nm pitch. (b) Depth-dependent Cram\'er--Rao bounds obtained with the archived 48$\times$48 ideal detector and identical detected-photon budget. The annular-vortex and fourfold-polynomial tests are screening calculations, not claims against globally optimized published double-helix or Tetrapod systems.}\label{fig:recovered}\label{fig:recovered_curves}
\end{figure}

The original geometry and complex coefficient lookup were recovered from the historical design archive. The archived array has 48$\times$48 locations with 1,716 active pillars, 113 distinct assigned diameters between 90 and 322 nm, an optimized residual-phase map, and a 117-entry complex RCWA lookup. The original code implements the detected-photon Poisson likelihood used to generate its historical numbers; it does not implement a calibrated physical camera or a displaced-dipole source. Figure~\ref{fig:recovered} displays the recovered physical assignment rather than a schematic substitute.

A second, NumPy-based Fourier-vector evaluator was written independently of the archived optimization script. It takes the saved diameter matrix, performs exact diameter-to-complex-coefficient lookup, removes the nominal focusing phase, computes the three electric field components and their \emph{analytical} position derivatives, and uses the five-parameter Poisson information matrix with the nuisance-variable Schur complement. The focal axial bound is 20.7695 nm, versus 20.7738 nm in the historical record (absolute discrepancy 0.0044 nm). Thus the \emph{local-periodic statistical prediction} can now be reproduced directly from the historical geometric dataset. It does not establish full-device Maxwell accuracy.

For a limited, matched comparison, the same evaluator was applied to a phase-only annular-vortex hologram family used to generate double-helix-type PSFs \cite{dhspiral}. The integer annulus count was selected by the mean positional log-determinant objective over five depths from $-1.5$ to $+1.5~\mu$m; all masks were mapped through the archived RCWA amplitude lookup and evaluated with the same 48$\times$48 detector and photon budget. Among the tested 2--12 annuli, the best member of that family yielded a focal axial bound of 23.876 nm. A separately optimized astigmatic family yielded 29.179 nm at focus (the optimization maximizes \emph{depth-averaged joint information}, not specifically focal $z$ information). As a diagnostic, a low-dimensional fourfold phase-polynomial proxy yielded 33.196 nm; it is \emph{not} an experimental or standard Tetrapod design and cannot be used to claim superiority to Tetrapod microscopy. The source archive reports 27.491 nm for its differently optimized astigmatic reference, which should not be conflated with the newly optimized family.

\begin{table}[t]
\centering\caption{\textbf{Focal localization screen with recovered original geometry.} All values are obtained from the same idealized 48$\times$48 Fourier detector, 3,000 detected photons and 150 background photons. The annular-vortex family search is finite; the fourfold proxy is explicitly not a published Tetrapod mask.}\label{tab:recovered}
\begin{tabular}{p{.60\linewidth}r}\toprule
Pupil design&$\sigma_z(0)$ (nm)\\\midrule
Recovered original Fisher-mapped metalens&20.769\\
Optimized annular-vortex DH family (2--12 rings)&23.876\\
Astigmatism, optimized for mean joint information&29.179\\
Fourfold polynomial proxy (not Tetrapod)&33.196\\\bottomrule
\end{tabular}
\end{table}

Finally, using the seven diameters for which a previously archived 81-, 121-, and 181-harmonic comparison is available, we interpolated their complex-transmission corrections across the recovered diameter map. The focal bound moved from 20.7695 to 20.7847 and 20.7863 nm in this \emph{conditional smooth-interpolation test}. Interpolation cannot detect narrow resonances between the sampled diameters, nor does the small change in a CRB imply convergence of the complex electromagnetic field. The complete 3D aperture scattering and vector dipole collection remain distinct unresolved computations. Exploratory full-aperture Yee--FDTD runs at coarse resolutions have also been attempted, but their fields failed spatial and temporal convergence and are excluded from the quantitative claims here. A high-resolution, open-boundary Meep/MPI calculation of the actual finite device is \emph{not available}.
\FloatBarrier

\section{Independent periodic-cell Maxwell check of the recovered map}\label{sec:independenthighorder}
The recovery of the actual diameter distribution allows a more direct test of the normal-incidence Maxwell lookup than comparisons based on a surrogate pupil. We recalculated three diameters used by the archived 48$\times$48 device (90, 230, and 322 nm) using an independent vector Fourier-modal solver with 289 retained spatial harmonics. These diameters occupy 8, 28, and 3 active sites, respectively. This calculation resolves the Maxwell eigenmodes of a periodically repeated cylindrical pillar cell, rather than the electromagnetic response of the \emph{finite} metalens. Its complex zeroth-order transmission coefficient is normalized consistently with the fused-silica outgoing medium. For all lossless periodic-cell calculations, the flux balance is verified independently of phase convergence.

\begin{table}[t]
\centering
\caption{\textbf{Direct independent Maxwell cell check.} The archived library uses 69 retained harmonics and substrate index $1.4585$; the independent results retain 225 and 289 harmonics with substrate index $1.45$. Inter-solver phase changes therefore combine differences in modeling assumptions and numerics, rather than isolating a truncation error.}\label{tab:independent289}
\begin{tabular}{rrrrr}\toprule
$d$ (nm)&Sites&$T_{289}$&$\phi_{289}$ (rad)&$\phi_{289}-\phi_\mathrm{archive}$ (rad)\\\midrule
90&8&0.978964&1.425218&$-0.05426$\\
230&28&0.969128&$-2.009902$&$-0.13350$\\
322&3&0.832994&1.388137&$-0.10191$\\\bottomrule
\end{tabular}
\end{table}

\begin{figure}[htbp]
\centering
\includegraphics[width=\linewidth]{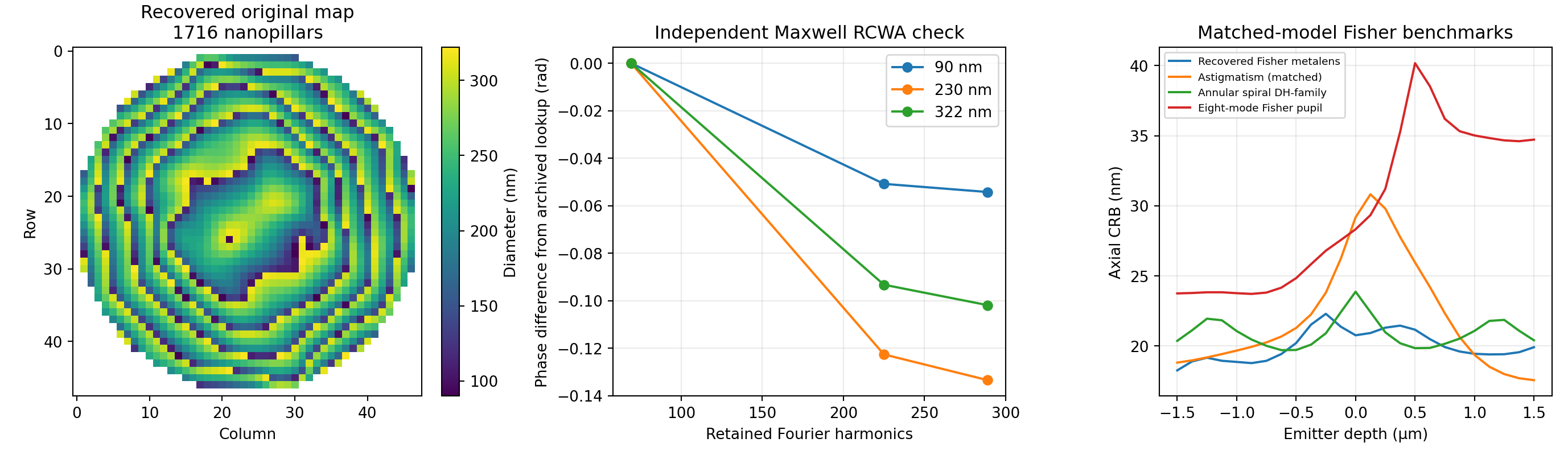}
\caption{\textbf{Recovered geometry and independent validation.} (Left) Site-resolved diameter map. (Middle) Deviation in zeroth-order transmission phase from the archived coefficient versus Fourier truncation for selected pillars; the lowest point is from the recovered archived coefficient library and the higher-order points were obtained from a separate vector RCWA implementation with a slightly different substrate index. (Right) Calculated axial local CRB under the archived, idealized, equal-detected-photon vectorial Poisson model. The flexible eight-mode pupil and annular-vortex double-helix family are restricted numerical screens, not experimental Tetrapod comparisons.}\label{fig:independentval}
\end{figure}

The 230-nm pillar is particularly informative. Its phase changes by $-0.01085$ rad between 225 and 289 harmonics, and differs by $-0.13350$ rad from the archived normal-incidence library. These discrepancies are not an uncertainty bound: they are comparisons of a finite number of discretizations. The archive and independent solver also use different substrate indices ($1.4585$ and $1.45$, respectively), so only the within-solver 225-to-289 harmonic difference isolates the effect of modal truncation under fixed material assumptions. Figure~\ref{fig:independentval} places this electromagnetic check alongside the original geometry and the matched-family localization benchmark.

To probe the impact of these differences without interpolating unsupported material response, we replaced \emph{only} the three corresponding diameter coefficients at their 39 actual sites and recomputed the complete pupil field and nuisance-aware Fisher matrix. The focal axial bound moved from 20.7695 to 20.7522 nm. This small change is a \emph{partial-map perturbation result}, not a sensitivity or convergence bound on the other 1,677 sites, and it cannot be used to certify the full-aperture Maxwell prediction.

We also tested a flexible eight-mode phase family optimized with three distinct initial conditions over the same five training depths and the same detected-photon model as the archived device. Its best local solution achieved a depth-averaged scaled log-determinant objective of 13.737, compared with 14.748 for the recovered Fisher map, and a focal axial bound of 28.33 nm. The screened annular-vortex double-helix family had a focal axial bound of 23.876 nm. These are bounded comparisons, not proof of global optimality of any pupil family, and no published physical Tetrapod design was reproduced. With the precise original diameter map now recovered, the remaining decisive experiment is a converged open-boundary \emph{finite-device} Maxwell calculation for displaced vector dipoles followed by detector-matched joint-localization tests.
\FloatBarrier

\section{Numerical diagnostics: modal response, phase mapping, and pixel integration}
\subsection{Periodically repeated pillar cells: power is easier to stabilize than phase}
The seven calculated diameters give transmitted powers between 0.8323 and 0.9854 at 225 harmonics (Table~\ref{tab:rcwa}). The largest lossless energy-balance residual is approximately $5.96\times10^{-13}$. Yet the 169-to-225 change in phase reaches 0.02827 rad for the 230-nm pillar. The 300-nm cell also retains a transmitted-power change of about 0.00209. These are differences between truncations, not certified upper bounds on the continuum error.

\begin{table}[t]
\centering\caption{Normal-incidence, lossless periodic-cell Maxwell/RCWA results at 225 Fourier harmonics. Phase differences and power differences are against the 169-harmonic computation. The numerical values are transcribed from an archived modal-convergence audit; the original CSV is not included with this preprint.}\label{tab:rcwa}
\begin{tabular}{rrrrr}\toprule
$d$ (nm)&$T_{225}$&$\arg t_{225}$ (rad)&$|\delta\phi|$ (rad)&$|\Delta T|$\\\midrule
90&0.97910&1.429&0.00490&0.000194\\
150&0.98545&2.152&0.00774&0.000074\\
200&0.96563&$-2.982$&0.01294&0.000011\\
230&0.96911&$-1.999$&0.02827&0.000059\\
260&0.97432&$-0.858$&0.01279&0.000043\\
300&0.90836&0.667&0.02138&0.002088\\
322&0.83232&1.397&0.01157&0.000818\\\bottomrule
\end{tabular}
\end{table}

\begin{figure}[t]
\centering\includegraphics[width=.96\linewidth]{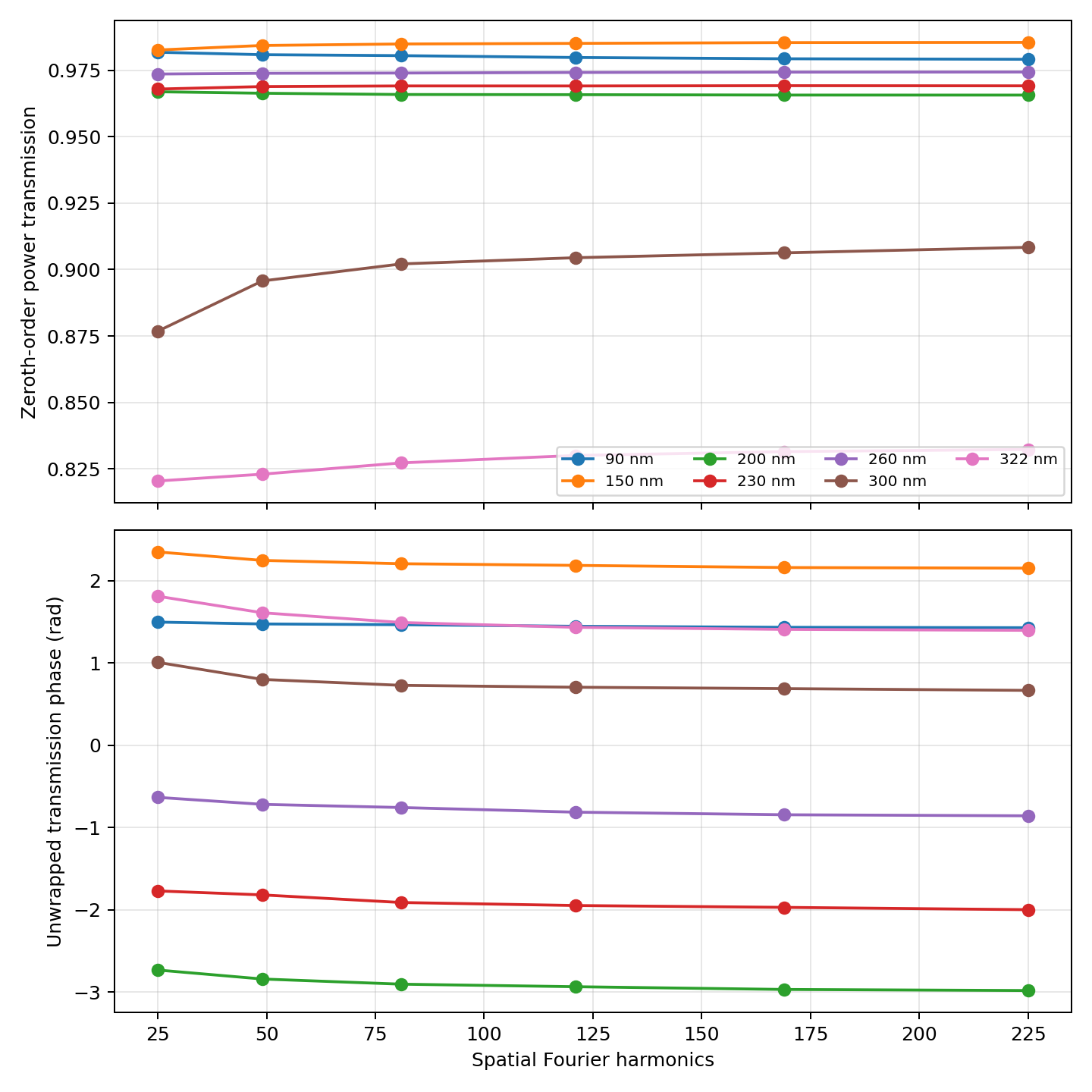}
\caption{\textbf{Vectorial periodic-cell RCWA.} Dependence of transmitted power and phase on Fourier-harmonic truncation for the tested TiO$_2$ geometries. The plotted energy balance does not alone establish convergence of complex transmission.}\label{fig:rcwa}
\end{figure}
\begin{figure}[t]
\centering\includegraphics[width=.90\linewidth]{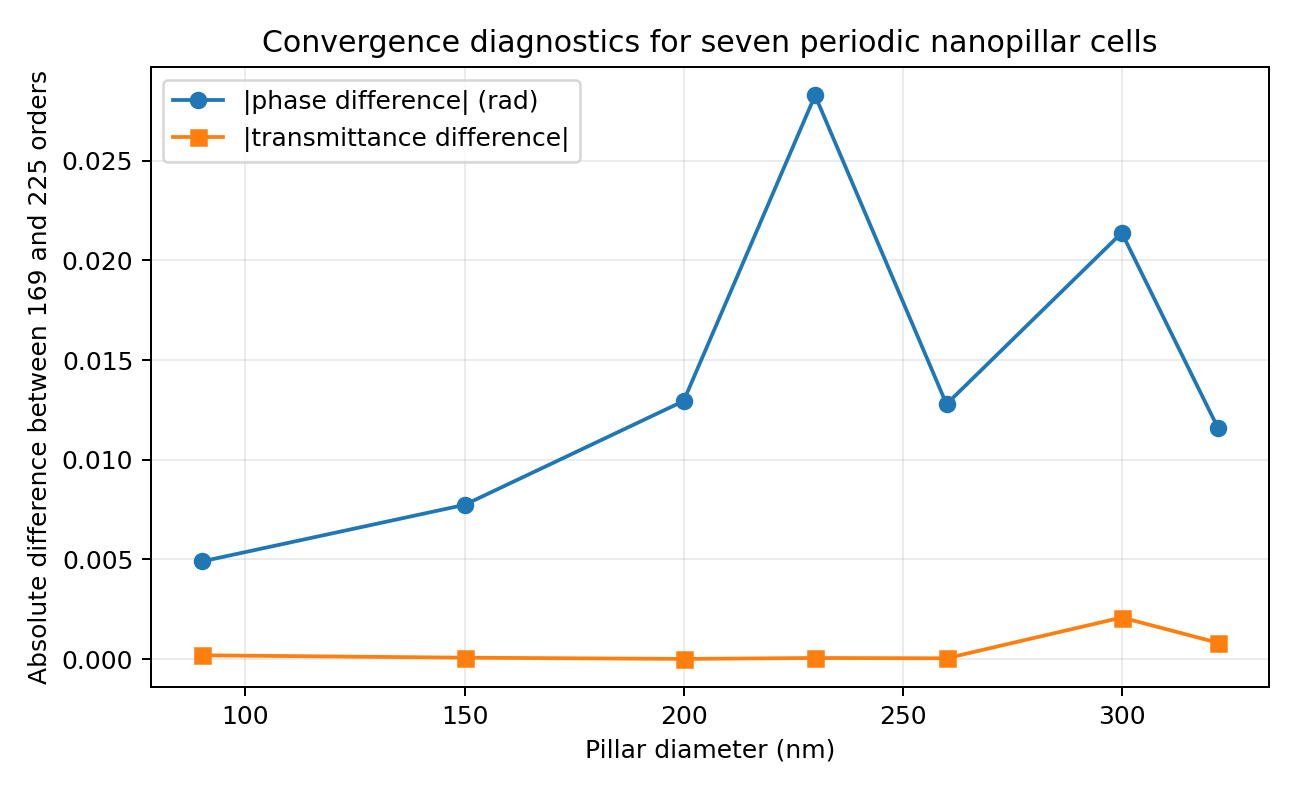}
\caption{\textbf{Complex-transmission audit.} Changes in periodic-cell power, phase, and complex coefficient between the two highest tested RCWA truncations. Both phase and complex-amplitude stability matter for PSF encoding.}\label{fig:complex}
\end{figure}

A phase-unwrapped traversal of the sampled diameters spans approximately 6.2513 rad, close to $2\pi$, but that global span should not be mistaken for fine phase addressability. The largest gap on the circular phase axis is 1.5247 rad. Assignment to the nearest of seven phases can therefore incur nearly half that gap as an angular quantization error. Moreover, two of the seven diameters satisfy the stringent provisional phase-difference check $|\delta\phi|<0.01$ rad. Those failures motivate a denser and demonstrably converged library, not interpolation of unverified phases.

\subsection{What seven available phases do to a test focusing pupil}
In a separate \emph{scalar} propagation diagnostic, 1741 sites were populated by the nearest available complex RCWA coefficient for a nominal hyperbolic phase. The resulting phase residual had root-mean-square magnitude 0.339 rad (95th percentile 0.636 rad). When each site is amplitude-matched to a lossless phase-only reference, the normalized focal coherence ratio is about 0.8905. Retaining the modeled transmitted amplitudes gives a relative on-axis intensity ratio of 0.8515 and a mean assigned cell transmission of 0.9563. These are ratios within the same scalar summation model, not absolute focusing efficiencies for a fabricated or finite Maxwell-simulated metalens.

\begin{figure}[t]
\centering\includegraphics[width=.93\linewidth]{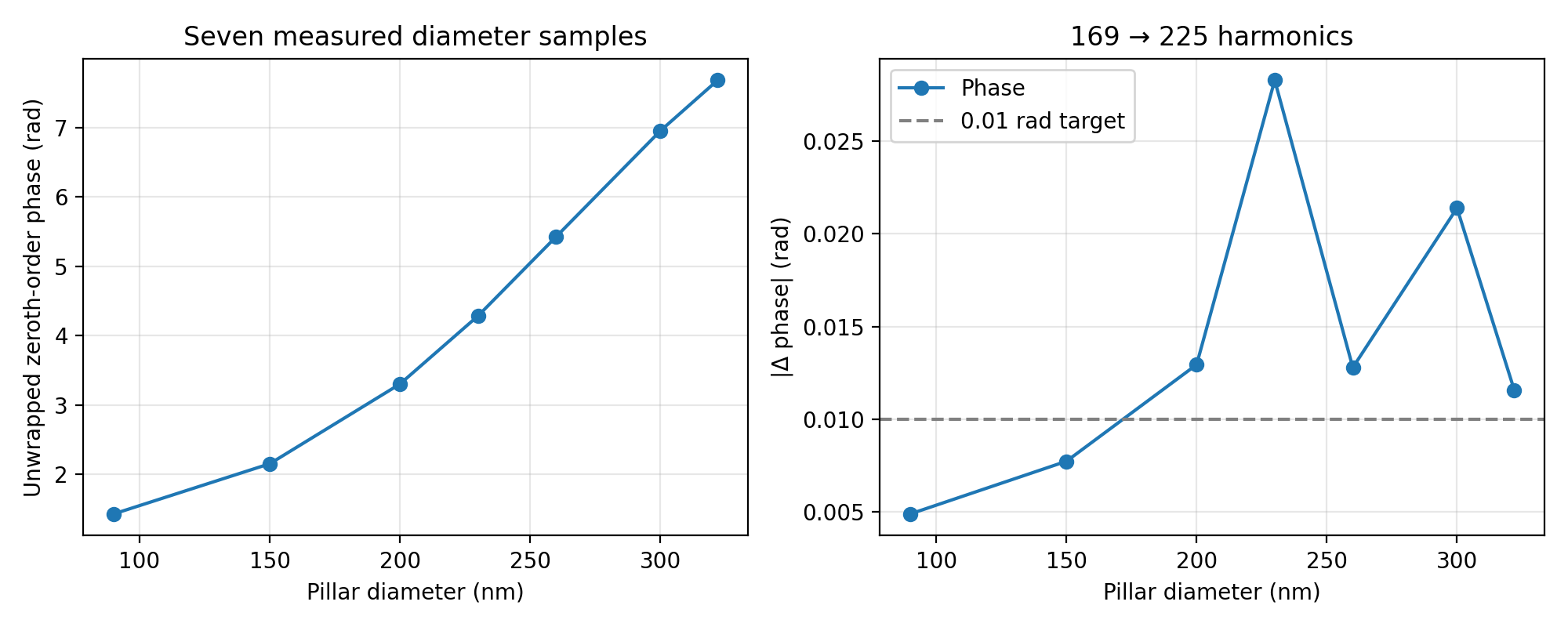}
\caption{\textbf{Phase coverage of seven computed pillar geometries.} Unwrapped phase progression and modal sensitivity reveal a sparse sampling gap and unresolved phase convergence.}\label{fig:phase}
\end{figure}
\begin{figure}[t]
\centering\includegraphics[width=.95\linewidth]{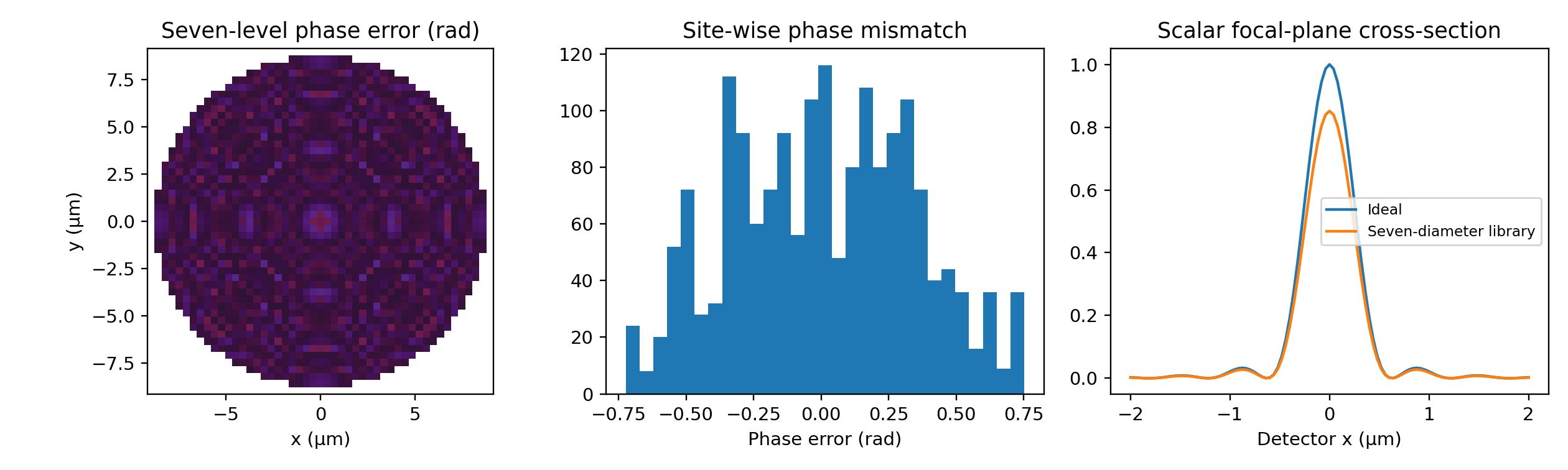}
\caption{\textbf{Scalar diagnostic of seven-phase quantization.} Spatial phase errors and normalized focal profiles for the representative hyperbolic pupil. This is not a vectorial Maxwell solution of a finite device.}\label{fig:quant}
\end{figure}

\subsection{The same RCWA lookup carried through vectorial imaging and pixel-integrated Fisher information}
The comparison in Table~\ref{tab:fisher} uses the same 25$\times$25 pixel detector, 0.20-$\mu$m pixel pitch, 3000 detected signal photons, and 150 background photons for both surrogate residual phases. With an 80$\times$80 pupil sample and 3$\times$3 Gauss--Legendre quadrature per pixel, the astigmatic pupil has a lower axial bound than the chosen mixed phase at the focus in each tested modal truncation. There is no claim of universal optimality: a different phase family or axial range can change the ordering.

\begin{table}[t]
\centering\caption{\textbf{Matched \emph{surrogate} Fisher comparison at focus.} RCWA cell response is locally assigned, but neither mask is the archived Fisher-optimized metalens or a fully optimized double-helix/Tetrapod comparator. Values are local bounds in nm, conditioned on 3000 detected signal photons.}\label{tab:fisher}
\begin{tabular}{rcc}\toprule
RCWA harmonics&Astigmatic $\sigma_z$&Mixed phase $\sigma_z$\\\midrule
121&20.844&21.618\\
169&20.878&21.561\\
225&20.862&21.549\\\bottomrule
\end{tabular}
\end{table}

The predicted astigmatic value changes by 0.077\% between 169 and 225 harmonics, whereas the underlying unit-cell phase can still differ by more than 0.02 rad. Stable output precision is therefore not a substitute for field convergence. The comparison of central sampling with area quadrature is likewise nontrivial: for the astigmatic 225-harmonic case, the pixel-center approximation yields 20.249 nm, while 2$\times$2 and 3$\times$3 quadratures give 20.86155 and 20.86165 nm. In an independent derivative check, the analytic $x$ derivative of the pixel-integrated intensity agreed with a central finite difference at a 0.5-nm source-displacement step to relative $L^2$ error $9.3\times10^{-7}$.

A further sampled-pupil check at 96, 112, and 128 points per axis gives astigmatic 225-harmonic bounds of 20.8068, 20.8067, and 20.8158 nm. The mixed-phase counterparts are 21.4426, 21.3634, and 21.3404 nm, retaining a greater residual dependence on pupil sampling. The normalized detector-PSF total-variation distances between 169 and 225 harmonics are approximately 0.0030 (astigmatic) and 0.00335 (mixed). The results show that the convergence of a normalized PSF, its derivatives, and the underlying electromagnetic field must be evaluated separately.

\begin{figure}[t]
\centering\includegraphics[width=.94\linewidth]{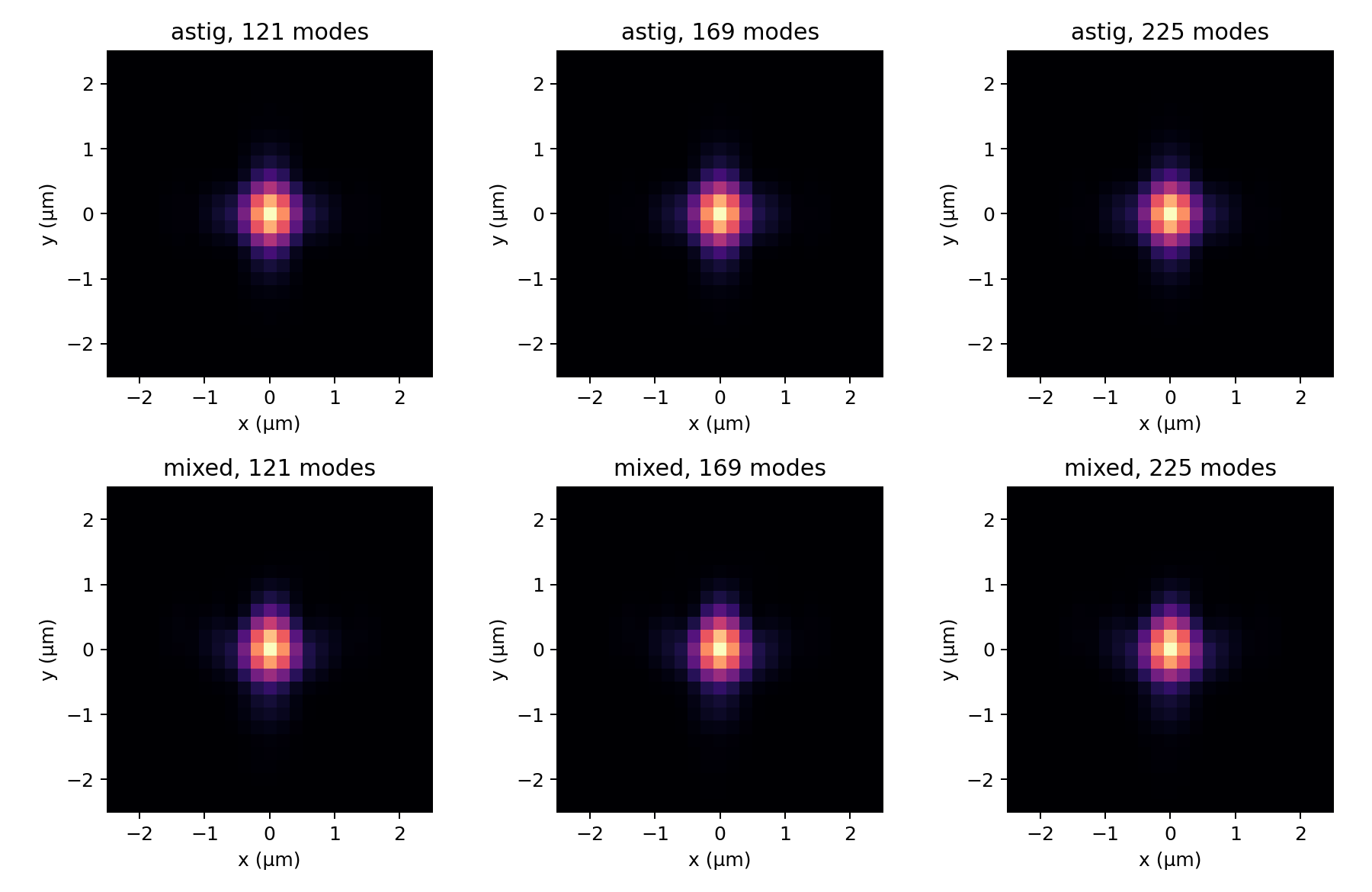}
\caption{\textbf{Calculated, not measured, detector PSFs.} Local-periodic RCWA coefficients are passed through the vectorial imaging model and integrated into 25$\times$25 pixels. Columns compare 121, 169, and 225 harmonics for the two illustrative phases.}\label{fig:psfs}
\end{figure}
\begin{figure}[t]
\centering\includegraphics[width=.93\linewidth]{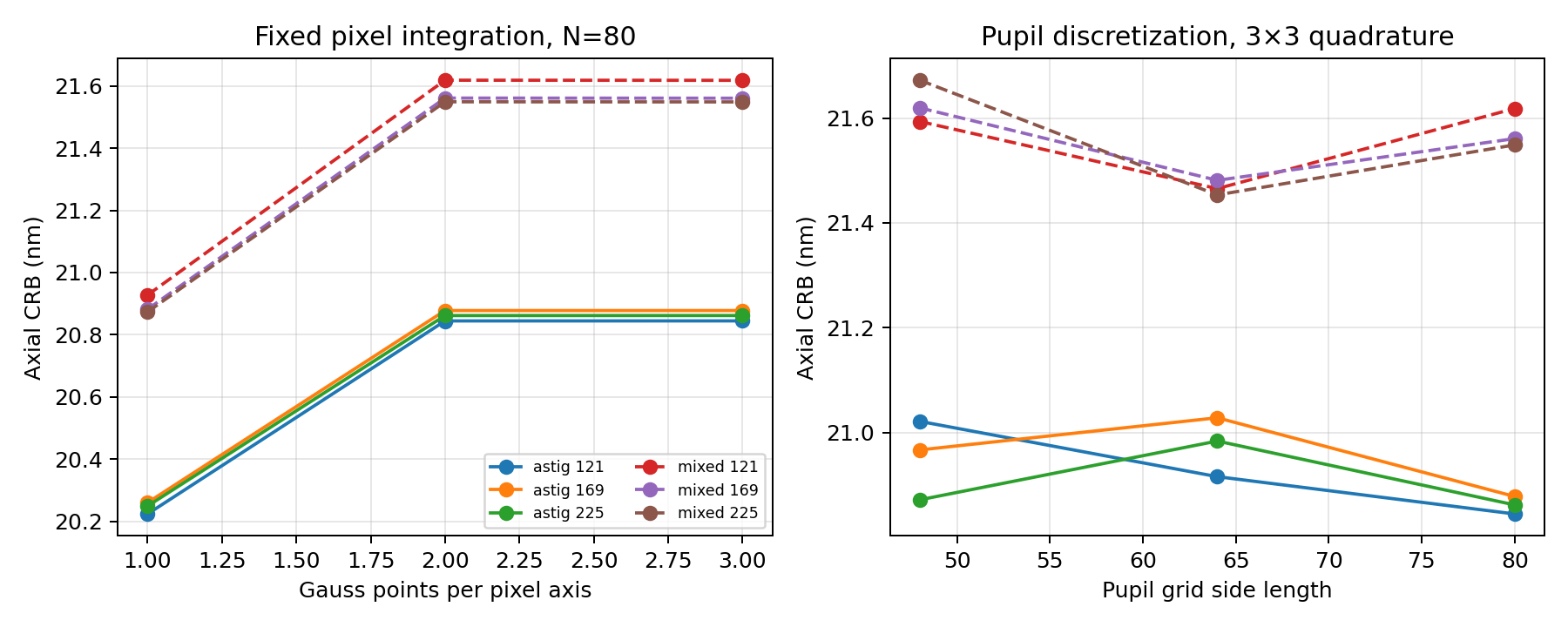}
\caption{\textbf{Numerical consistency of surrogate Fisher predictions.} Dependence of the axial bound on pixel quadrature and sampled-pupil resolution at a fixed physical detector and identical detected-photon budgets. Stable bounds do not establish finite-metalens Maxwell convergence.}\label{fig:fisherplot}
\end{figure}
\FloatBarrier

\section{Discussion: evidence, novelty, and the missing device test}
The principal result is a reproducible, geometry-specific local-periodic information benchmark, accompanied by a quantitative demonstration of what a sparse or incompletely converged electromagnetic library can and cannot support. The meaningful outcome is not that an arbitrary phase mask outperforms another. The two illustrative families are deliberately simple, and their relative ranking is not a claim against established Fisher-optimized phase masks \cite{shechtman2014}, double-helix localization \cite{wang2023}, or experimentally calibrated PSF models \cite{xiao2025}. The information gained here is more specific: a traceable local-periodic implementation may show well-behaved output intensity and Fisher bounds while some inputs---notably the cell phase and diameter identity---have not converged. The result provides a concrete warning against inferring device-level metrological superiority from focal power or from a single numerical precision estimate.

Several different checks support that inference. First, the homogeneous-film Fresnel test and lossless power balance establish internal consistency of the modal implementation in those cases. The archived and independent solver results also differ slightly in substrate-index assumptions; their phase separation must not be interpreted as an isolated truncation error. Second, explicit differences between Fourier truncations expose phase errors that energy conservation cannot reveal. Third, nearest-neighbor phase assignment converts those differences into discontinuous, site-dependent changes. Finally, direct differentiation and finite-area pixel integration show how to calculate the local statistical bound without conflating a digital image shift with a physically displaced emitter. Similar insistence on comparing a forward model against a defined physical quantity is evident in optical-design work by Moreno and collaborators \cite{moreno2021}; however, their LED irradiance and freeform-surface testing settings are distinct from the present meta-optical localization problem \cite{delolmo2021}.

The historical optimized metalens and its site-resolved geometry can now be reproduced under the original 48$\times$48 statistical model (Sec.~\ref{sec:recovered}). That computation, however, differs from the independent pixel-integrated 25$\times$25 surrogate model above; directly pooling their bounds would still be misleading. A phase-mask-family screen is not a claim of global superiority over experimental double-helix or Tetrapod systems. The local-periodic coefficients also omit angle-dependent and polarization-dependent response away from the normal-incidence cell problem, as well as inter-pillar and edge coupling in a finite aperture. These effects may alter the PSF in ways that the present stability analysis cannot bound.

\begin{table}[t]
\centering\caption{Evidence status and the reproducible requirements for a future device-level claim. Acceptance tolerances are \emph{proposed targets}, not thresholds already met.}\label{tab:gates}
\begin{tabular}{p{.36\linewidth}p{.49\linewidth}}\toprule
Observable&Status / required test\\\midrule
Unit-cell energy balance&Passes for lossless, normally incident cells; not a phase test.\\
Unit-cell complex phase&Not converged to proposed 0.01-rad change at five of seven diameters.\\
Phase addressability&Seven geometries are too sparse for unrestricted accurate mapping.\\
Detector Fisher calculation&Internally cross-checked for the illustrative local-periodic model.\\
Original optimized geometry&Recovered; local-periodic CRB independently reproduced.\\
Finite-array Maxwell and collection&No converged open-boundary vector solution.\\
Competitive PSFs&Annular-vortex DH family screened; actual Tetrapod and joint Monte Carlo still required.\\\bottomrule
\end{tabular}
\end{table}

The experiment or simulation that would settle the original physical-performance question is well defined. The recovered original diameter map can now be used to construct a sufficiently dense, angle- and polarization-resolved Maxwell library; propagate or simulate the actual finite device with verified open boundaries; and form a PSF for a displaced vector dipole at fixed detector geometry. Its Jacobian must be converged along with the intensity itself. A fair incident-photon-budget comparison should use independently optimized astigmatic, double-helix, and Tetrapod references, identical detection parameters, and joint maximum-likelihood trials reporting bias, outliers, and RMSE as well as CRBs. Those definitive full-device and incident-photon comparisons have not been performed here, and the present paper does not claim otherwise. Coarse preliminary full-aperture FDTD tests do not converge and are not used as supporting performance evidence.

\section{Conclusions}
We reconstructed an archived site-resolved 1,716-pillar dielectric-metalens model and independently reproduced its local axial Fisher bound: 20.7695 nm versus 20.7738 nm in the historical record, conditioned on the same 48$\times$48 ideal Fourier detector and a fixed detected-photon budget. The narrow numerical agreement certifies reproducibility of \emph{that particular RCWA/local-periodic forward and noise model}, not the precision of a fabricated or Maxwell-resolved finite metalens.

Independent vector Fourier-modal tests for three assigned nanopillar diameters expose non-negligible deviations in complex transmission. The 230-nm phase differs by 0.1335 rad from the archive, a comparison that mixes numerical and modest substrate-index differences and therefore is not a truncation-only error estimate. Replacing coefficients at just 39 known sites moves the bound to 20.7522 nm, illustrating a well-defined but highly localized response test rather than a full-device uncertainty bound. Seven-diameter surrogate checks further show that a numerically stable Fisher bound can coexist with unresolved cell phases and nontrivial phase quantization.

Within a common idealized detector and detected-photon convention, the recovered design also has a lower focal axial bound than the restricted annular-vortex and astigmatic families screened here; the finite screen does not establish superiority over optimized, calibrated double-helix or Tetrapod systems. The transferable conclusion is methodological: the phase library, pixel-resolved PSF derivatives, statistical nuisance treatment, and actual finite-device response must be audited separately before an information-optimized pupil can be claimed to achieve a physically realized localization precision.

\section*{Research scope and data provenance}
This preprint reports an independent reproduction and numerical audit of a previously constructed RCWA/local-periodic metalens model. The reported 20.7695-nm localization bound is conditional on its idealized forward and detector models and is not an experimental precision or a converged finite-aperture three-dimensional Maxwell result. Source calculations and detailed procedures are described in Appendix A. The full historical design archive and numerical analysis scripts are not distributed in this arXiv source bundle; readers should not interpret the present article as an independently executable data repository. Questions about numerical inputs and availability can be addressed to the corresponding author. Both listed authors should verify the manuscript and data provenance before public release.

\appendix
\section{Supplementary numerical methods and scope}
\subsection{Provenance of reported simulations}
The main article is an analysis of existing numerical results. The periodic-cell vector RCWA coefficients originated in the numerical working archive, which is not distributed with this preprint. A later audit recovered the historical original site-resolved geometry and the 117-entry RCWA transmission lookup. The recovered diameter and residual phase matrices and independent evaluator were inspected in the working archive but are not packaged with the present article. These are distinct from the seven-diameter surrogate dataset. Local-periodic benchmark values are reported in the manuscript; numerical code and underlying CSV tables are not included in this manuscript-only submission.

\subsection{Reproduction of the recovered original geometry}
The historical archive \texttt{Metalens\_Information\_Optimized\_OE\_v12.zip} contains the 48$\times$48 physical design grid, 1,716 nonzero sites and 113 unique diameters. It also contains 117 complex periodic-cell RCWA coefficients. An independent NumPy script assigns the saved complex coefficients at exactly matching diameters, accounts for the nominal focusing phase and calculates the vectorial pixelated PSF and its source-coordinate derivatives. With the historical ideal Fourier detector, $N_s=3000$, and $b=150$, the resulting bound is $\sigma_z=20.76948$ nm at nominal focus; the historical report lists 20.77384 nm. This close internal replication is not an independent three-dimensional Maxwell finite-array solution.

An annular-vortex mask family, $P_N(\rho,\theta)=(2n-1)\theta$ on radial annulus $\sqrt{(n-1)/N}<\rho\le\sqrt{n/N}$, was screened at integer $N=2,\ldots,12$ in the same recovered detector model. The selected family member yielded $\sigma_z(0)=23.876$ nm. A limited fourfold polynomial screen is named a proxy because no authentic published Tetrapod phase mask was available for matched evaluation. A separate interpolated modal sensitivity correction, based on seven sample diameters, changes the original bound to $20.7863$ nm at a nominal 181-harmonic correction. This interpolation is conditional and cannot bound unknown diameter resonances or nonlocal coupling.

\subsection{Unit-cell Maxwell checks}
Periodic cells were evaluated at vacuum wavelength 850 nm, pillar height 1.0 $\mu$m, and square period 380 nm with refractive indices $n_{\mathrm{TiO_2}}=2.4116$ and $n_{\mathrm{sub}}=1.45$, assumed dispersionless and lossless for this monochromatic study. The archived original map uses $n_{\mathrm{sub}}=1.4585$ in its coefficient normalization, so archived-to-new solver differences should not be attributed solely to truncation. Reported harmonic counts are the total retained transverse Fourier orders. The seven diameters are 90, 150, 200, 230, 260, 300, and 322 nm. Power balance for these lossless unit cells satisfies
\[
\delta_{\mathrm{balance}}=|T+R-1|\lesssim6\times 10^{-13}
\]
at the last stored modal truncation. This test alone does not bound the phase error. Convergence is audited with $\arg(t_{225}t^*_{169})$ and the absolute change in normalized complex transmission. The largest phase discrepancy is about 0.02827 rad for 230 nm, well above a stringent provisional 0.01-rad adjacent-truncation target. Therefore the study explicitly reports a partial, not certified, complex-library convergence.

\subsection{Finite-geometry and source-model limitations}
The periodic RCWA model assumes all adjacent nanopillars share one diameter and that illumination is normally incident. Its coefficients are used in a nearest-phase local-periodic mapping of a finite aperture, where high-angle incidence, neighbor coupling, edges, and the finite substrate are not treated rigorously. A direct source-scattering calculation would require a displaced vector electric dipole and a three-dimensional electromagnetic solver with correctly tested open-boundary absorbers. The optimized diameter map has now been recovered, but no converged full-device Maxwell scattering solution is available. Exploratory full-aperture Yee--FDTD computations at coarse spatial resolutions were performed, but showed substantial grid and time dependence; they are excluded from the scientific evidence supporting this manuscript. High-resolution, open-boundary Meep/MPI simulations have not been executed.

\subsection{Forward model and camera discretization}
The fields use a transverse angular-spectrum approximation with a fixed circular pupil, x-polarized illumination, and a Debye-like apodization. Spatial derivatives are evaluated analytically under the angular-spectrum integral. Each camera pixel is assigned finite physical area and integrated with tensor-product Gauss--Legendre quadrature. At $N_s=3000$ detected signal and $b=150$ spatially uniform background counts, the means are
\[
\mu_j=N_s p_j(x,y,z)+b/M.
\]
The full five-parameter Poisson Fisher matrix is evaluated analytically from pixel-integrated intensity derivatives. The positional Fisher matrix is the Schur complement over $N_s$ and $b$; reported CRB entries are square roots of covariance lower bounds expressed in nm. Treating $N_s$ as the \emph{detected} budget deliberately removes a difference in absolute throughput; a matched \emph{incident}-photon comparison cannot be claimed from these calculations.

\subsection{Checks on derivatives and quadrature}
For a local-periodic astigmatic test, a centered difference with source step 0.5 nm was compared to the analytic derivative of the pixel-integrated intensity: the reported relative Euclidean norm discrepancy was approximately $9.3\times10^{-7}$. With $80\times80$ pupil points and a 25$\times$25 camera, the astigmatic 225-harmonic axial bound changes from 20.249 nm (center sampling) to 20.86155 nm (2$\times$2 quadrature) and 20.86165 nm (3$\times$3 quadrature). This supports quadrature stability within the stated \emph{surrogate}; it does not imply convergence of the physical meta-atom response. For grid sizes 96, 112, and 128 at 225 harmonics, the astigmatic CRB values are 20.8068, 20.8067, and 20.8158 nm, respectively; mixed residual phases show larger remaining grid variation.

\subsection{Quantization test and its interpretation}
The seven-coefficient lookup was also used in an independent scalar focusing calculation with 1741 sites. Nearest-phase assignment gives RMS phase error about 0.339 rad, 95th-percentile absolute error 0.636 rad, and a normalized relative focal-peak ratio 0.8515 when amplitude losses within the lookup are retained. This is neither a vectorially computed focusing efficiency nor a comparison to the original Fisher-optimized device. It is presented as a diagnostic of the sparse library's inability to approximate an arbitrary continuous phase without substantial quantization.

\subsection{Publication-level acceptance criteria not yet satisfied}
\begin{enumerate}
\item Recover or independently re-optimize and distribute the \emph{exact} site-by-site geometry.
\item Produce a dense complex RCWA library and check dependence on harmonic count, angle, polarization, finite material dispersion, and fabrication perturbations.
\item Compute a finite-pillar-array Maxwell solution with verified open-boundary convergence and at least two independent grid levels.
\item Propagate a physically specified displaced-dipole response through the actual imaging train and integrate a calibrated finite-pixel detector.
\item Compare against equally optimized astigmatic, double-helix, and Tetrapod baselines at fixed incident photon flux, and report Fisher bounds, estimator bias, RMSE, and failure rate.
\end{enumerate}
These are requirements for a \emph{future claim} of finite-device superiority. Listing them does not imply that the tests were executed or passed.

\subsection{Reproducibility limitations of this preprint package}
The present arXiv source contains the complete manuscript, figures, and numerical methods narrative, but not the historical input geometry, solver code, or validation outputs as a reusable computational archive. Exact independent numerical regeneration therefore requires access to the underlying dataset and scripts. In particular, the missing finite-device Maxwell run must not be inferred from the reported RCWA and Fourier--vector surrogate calculations.

\end{document}